\documentclass[12pt,a4paper]{iopart}

\newcommand{\eq}[1]{\begin{equation}#1\end{equation}}
\newcommand{\dd}{\mathrm{d}}

\newcommand{\ch}{\mathrm{ch}}

\newcommand{\Ai}{\mathrm{Ai}}
\usepackage{iopams}
\usepackage{graphics}
\usepackage{graphicx}
\usepackage{psfrag}

\begin{document}

\title{Surface and bulk entanglement in free-fermion chains}
\author{ Viktor Eisler$^1$ and Ingo Peschel$^2$}
\address{
$^1$Institute for Theoretical Physics, E\"otv\"os Lor\'and University,\\
P\'azm\'any P\'eter s\'et\'any 1/a, 
H-1117 Budapest, Hungary\\
$^2$Fachbereich Physik, Freie Universit\"at Berlin,\\
Arnimallee 14, D-14195 Berlin, Germany
}
\eads{\mailto{eisler@general.elte.hu} and \mailto{peschel@physik.fu-berlin.de}}

\begin{abstract}

We consider free-fermion chains where full and empty parts are connected by a 
transition region with narrow surfaces. This can be caused by a linear potential or by 
time evolution from a step-like initial state. Entanglement spectra, entanglement 
entropies and fluctuations are determined for subsystems either in the surface region or 
extending into the bulk. In all cases there is logarithmic behaviour in the subsystem 
size, but the prefactors in the surface differ from those in the bulk by $3/2$. A
previous fluctuation result is corrected and a general scaling formula is inferred 
from the data. 

\end{abstract}

\section{Introduction}

  The ground-state entanglement in critical quantum chains has been the topic of many studies, see e.g.
\cite{CC09}. For a subsystem of length $L$ in a larger total system, the entanglement entropy $S$ varies 
logarithmically, $S =c\nu/6 \,\ln L$, where $c$ is the central charge of the corresponding conformal 
field theory and $\nu= 1,2$ is the number of contact points. With boundaries, the location of the 
subsystem enters, but only via a finite additive term. While found originally for homogeneous chains, 
these features persist if one adds couplings or potentials which vary in space according to power laws.
Such studies have been performed
for XX spin chains \cite{Eisler/Igloi/Peschel09,CV10a,CV10b,CV10c}, 
the transverse Ising model \cite{Eisler/Igloi/Peschel09} 
the XY model \cite{CV10a,CV10c} and 
free fermions in the continuum \cite{Calabrese/Mintchev/Vicari12a,Vicari12}. In all these cases, one has 
logarithmic laws, either in the subsystem size or in a typical length connected with the perturbation 
or in the particle number, and the prefactor is the same as in the homogeneous case. However, 
although bipartitions of various kinds have been considered, one has been looking essentially only at 
bulk properties.

On the other hand, a system of free particles in a trapping potential has not only a varying density, but
also a well-defined surface region, where this density falls to zero smoothly. The same is true for 
fronts evolving from an initially step-like magnetization in XX \cite{ARRS99,HRS04} or XXZ chains 
\cite{Sabetta/Misguich13}, and it was pointed out recently that these surfaces have common universal 
features \cite{Eisler/Racz13,Eisler13}. In traps this is related to the linear variation of the potential 
near the Fermi energy \cite{Eisler13}, while the XX fronts can be mapped exactly to a static problem with 
a completely linear potential \cite{Eisler/Racz13}. This raises the question about the entanglement in such 
surfaces and its relation to the bulk properties. Some results have already been shown in 
\cite{Eisler/Racz13,Eisler13}. In this paper, we investigate the problem in more detail.

The system we study is a free-fermion chain in a linear potential \cite{Eisler/Igloi/Peschel09}. 
The ground state then has a sandwich structure with a full and an empty region connected by an interface 
of width $2\ell$. This interface terminates on both sides in surface regions with another typical length 
scale $\ell_s \ll \ell$. We consider subsystems located fully or partly in this surface region and 
determine their entanglement with the remainder. This is done numerically, using the correlation matrix 
and calculating its eigenvalues and eigenfunctions. These are qualitatively very similar to the bulk case.
As a result, one has logarithmic laws also in the surface region, both for the entanglement entropy and 
for the closely related particle fluctuations. However, the prefactors differ by $3/2$ from  
those in the bulk. For subsystems extending into the bulk, we also find a simple scaling form for $S$ 
which connects both regimes and also describes the case of a harmonic trap.

In our calculations we are actually treating the so-called Airy kernel, to which the correlation 
matrix reduces in the surface. This operator is well known in random-matrix theory \cite{Mehta04} where
it also describes an edge problem and where analytical results for the fluctuations exist 
\cite{Soshnikov00}. They confirm the logarithmic behaviour and indicate why the factor $3/2$ appears, 
but contain small errors which we locate and correct.

In the following, we first introduce our model and collect some facts on the procedure and on
the correlation matrix in section 2. Then, in section 3, we discuss its eigenvalues and 
eigenfunctions in the surface case. In section 4, we show the resulting entanglement entropy 
and fluctuations for various choices of the subsystem and compare with analytical results. 
In section 5, we analyze eigenvalues and entropy for the bulk region. A summary and discussion is 
given in section 6 and the appendix contains an outline of the correct fluctuation calculation.

\section{Setting and basic formulae}
 
 We consider a chain of free fermions with nearest-neighbour hopping in a linear potential
\cite{Eisler/Igloi/Peschel09}.
In spin language, this corresponds to an XX model with a gradient in the magnetic $z$-field.
The Hamiltonian is, for a finite system with open ends, 
\eq{
H= -\frac 1 2 \sum_n (c^{\dag}_n c_{n+1} + c^{\dag}_{n+1} c_{n})
+ \sum_n h\,(n-1/2) c^{\dag}_n c_{n} \, .
\label{hamXX}}
Here the linear term is chosen such that it goes through zero between
sites 0 and 1. This Hamiltonian has a single-particle spectrum $\omega_k$ with equidistant levels in 
the center forming the famous Wannier-Stark ladder. For a large system, only these
levels with $\omega_k=h(k-1/2)$ and eigenfunctions $\psi_k(n)=J_{n-k}(1/h)$ are relevant. 
Here the $J_n$ are the usual Bessel functions and their argument defines a characteristic 
length $\ell=1/h$, which is also the inverse slope of the potential.
In the ground state, the levels with $k \le 0$ are occupied and the density has the behaviour
shown in Fig. 1 of \cite{Eisler/Igloi/Peschel09}. Between a completely full region 
$n \lesssim -\ell$ and a completely empty one $n \gtrsim \ell$ there is an interface with a 
slow density variation in the center and a rapid one near the edges, where the potential 
reaches the critical values $\pm1$. Our main interest will be in this surface region, $n \sim \ell$, 
for large values of $\ell$.

The entanglement properties are determined by the correlation matrix 
$C_{mn}=\langle c^{\dag}_m c_{n} \rangle$ which is given by 
\eq{
C_{mn} = \frac{\ell}{2(m-n)} \left[J_{m-1}(\ell) J_{n}(\ell) - 
          J_{m}(\ell) J_{n-1}(\ell) \right] .
\label{corr}}
Restricting $C_{mn}$ to the chosen subsystem, its eigenvalues $\zeta_k$ give the single-particle 
eigenvalues $\varepsilon_k = \ln [({1-\zeta_k})/{\zeta_k}]$
of the entanglement Hamiltonian $\mathcal{H}$ in
\eq{ 
\rho = \frac{1}{Z} \exp(-\mathcal{H})
\label{rho}}
where $\rho$ is the reduced density matrix \cite{Peschel03,Cheong/Henley04}. 
From them, the entanglement entropy $S= -\mathrm{Tr}\,(\rho \ln \rho)$
follows as
\begin{equation}
S =  \sum_k \ln (1 + \mathrm{e}^{-\varepsilon_k})+\sum_k
\frac{\varepsilon_k}{\mathrm{e}^{\varepsilon_k} +1}
\label{entropy}
\end{equation}
and the particle-number fluctuations $\kappa_2 = \langle N^2 \rangle -\langle N\rangle^2$ in the 
subsystem as 
\begin{equation}
\kappa_2 =  \sum_k \zeta_k(1-\zeta_k)   
         =   \sum_k \frac {1}{4\,\ch^2(\varepsilon_k/2)} \, .
\label{fluct}
\end{equation}

The same density profile and correlation matrix as above also appear in a dynamical problem,
namely if one considers a chain which at time $t=0$ is completely filled for $n \le 0 $ and 
completely empty for $n \ge 1$ \cite{ARRS99}. Evolving this state with the Hamiltonian (\ref{hamXX}) 
with $h=0$, the correlation matrix at a later time $t$ is then given by (\ref{corr}) with 
$\ell$ replaced by $t$ and an additional factor $i^{m-n}$ which, however, does not affect the 
eigenvalues \cite{Peschel/Eisler09}. In this case, one is dealing with a moving front instead 
of a static surface.

The expression (\ref{corr}) simplifies in two limits. If $m$ and $n$ are kept finite and
$\ell$ is large, one is in the bulk of a wide interface and obtains 
\eq{
C_{mn} = \frac{\sin(\pi(m-n)/2)}{\pi(m-n)} 
\label{sine}}
which is the result for a half-filled homogeneous system.

If, on the other hand, $m=\ell+\ell_s \,x$ and
$n=\ell+\ell_s \,y$, one is within a certain range given by the length
\eq{ 
\ell_s=(\ell/2)^{1/3}
\label{surface_length}}
near the surface and one obtains for large $\ell$ \cite{Eisler/Racz13} 
\eq{
C_{mn} =  (1/\ell_s)\, K(x,y) 
\label{airy1}}
where
\eq{
K(x,y)=\frac{\Ai(x)\Ai'(y)-\Ai'(x)\Ai(y)}{x-y}
\label{airy2}
}
with the Airy function $\Ai(x)$. This so-called Airy kernel also appears, if one considers 
the surface region of a system of fermions trapped in a potential well \cite{Eisler13}.
Moreover, it has been known for quite some time in the theory of random matrices 
where it governs the statistics of the eigenvalues in the Gaussian unitary ensemble at the edge of 
the spectrum \cite{Mehta04}. All these surface problems are therefore connected. The surface length
$\ell_s$ can also be obtained from a scaling argument \cite{CV10c}.

In the following, we consider subsystems either in the surface region or reaching into the bulk 
and diagonalize the correlation matrix $C_{mn}$ numerically.

\section{Surface region: eigenvalues and eigenfunctions}

In this section, we consider subsystems extending from some point inside the surface all the way
into the empty region. Thus the variables $x$ and $y$ are taken in the interval $(-s,\infty)$, 
where $s>0$, and the eigenvalue problem of $C_{mn}$ is solved for different $s$. This is done by 
restricting the indices to $m_{min}< m,n <m_{max}$ where $m_{min} = \ell - \ell_s s$ and 
$m_{max}$ is chosen sufficiently large such that all matrix elements with $m,n \ge m_{max}$ 
are negligibly small. 

\begin{figure}[htb]
\center
\includegraphics[scale=0.7]{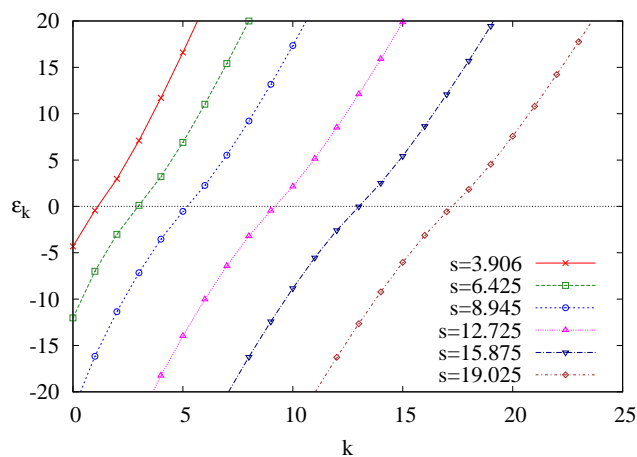}
\caption{Single-particle eigenvalues $\varepsilon_k$ of the entanglement Hamiltonian $\mathcal{H}$ 
for $\ell=1000$ and several values of $s$.}
\label{fig:eps}
\end{figure}

Fig. \ref{fig:eps} shows the resulting $\varepsilon_k$ for $\ell=1000$ ($\ell_s \simeq 8$) and 
several values of $s$ corresponding to integer values of $m_{min}$. The qualitative behaviour
is very reminiscent of that for homogeneous systems, namely one has a linear variation with
additional slight curvature \cite{Peschel/Eisler09,Eisler/Peschel13}. The leftmost spectrum 
refers to a subsystem with rather small filling, and there are only two negative $\varepsilon_k$ 
(corresponding to occupation numbers $\zeta_k > 1/2$). As the subsystem extends more into the 
surface region, the filling also increases, more negative $\varepsilon_k$ appear and the dispersion 
curve is pulled down successively. At the same time, the slope becomes smaller, which leads to an 
increase of the entanglement (see Section 4).
 
%
%

\begin{figure}[t]
\center
\includegraphics[width=0.3\textwidth]{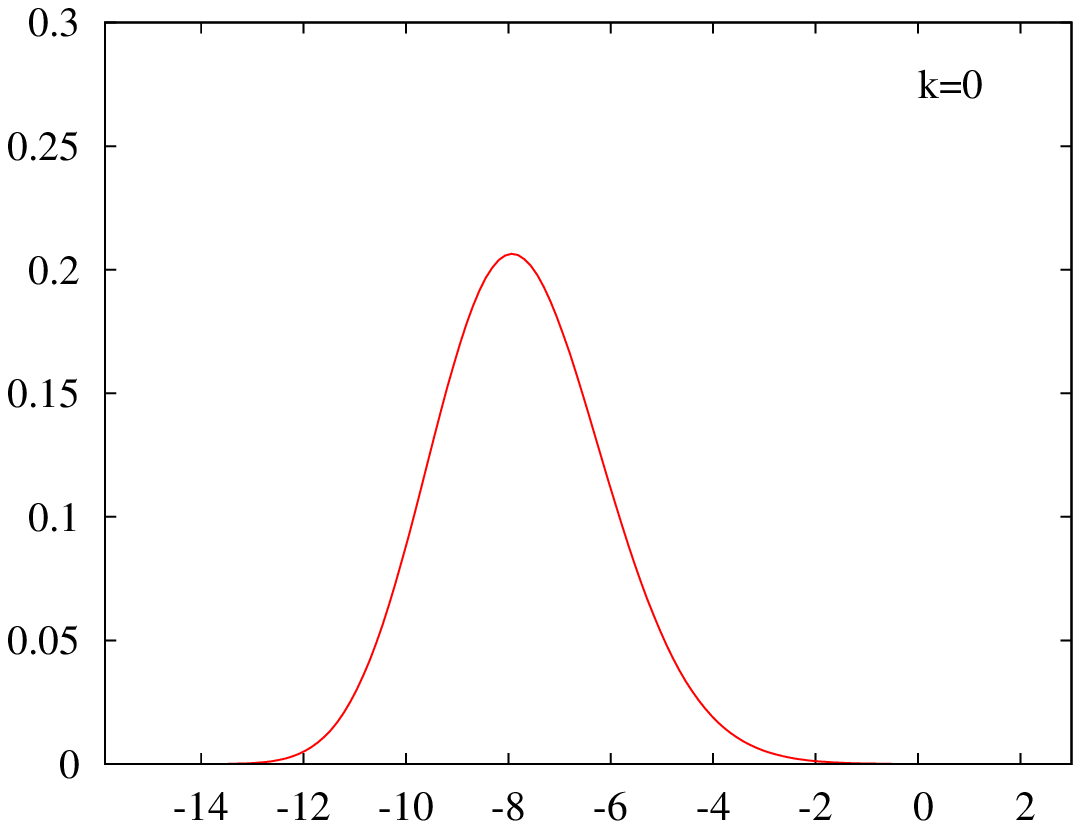}
\includegraphics[width=0.3\textwidth]{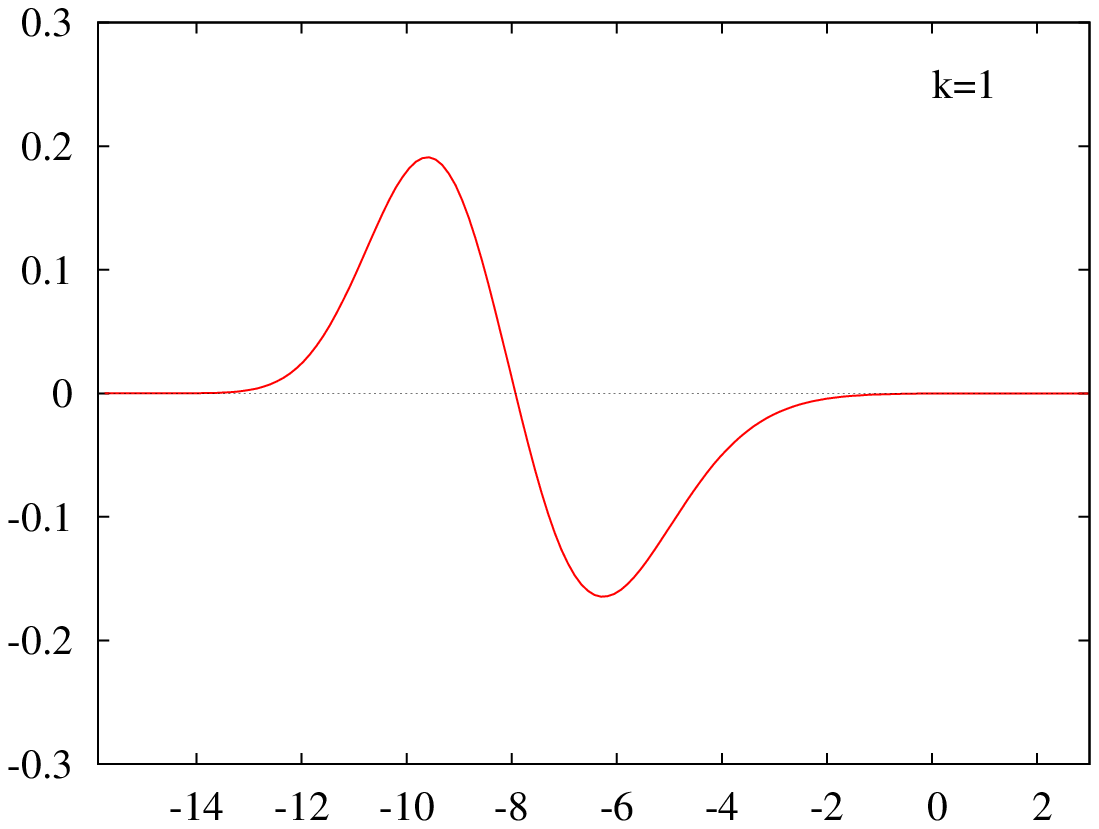}
\includegraphics[width=0.3\textwidth]{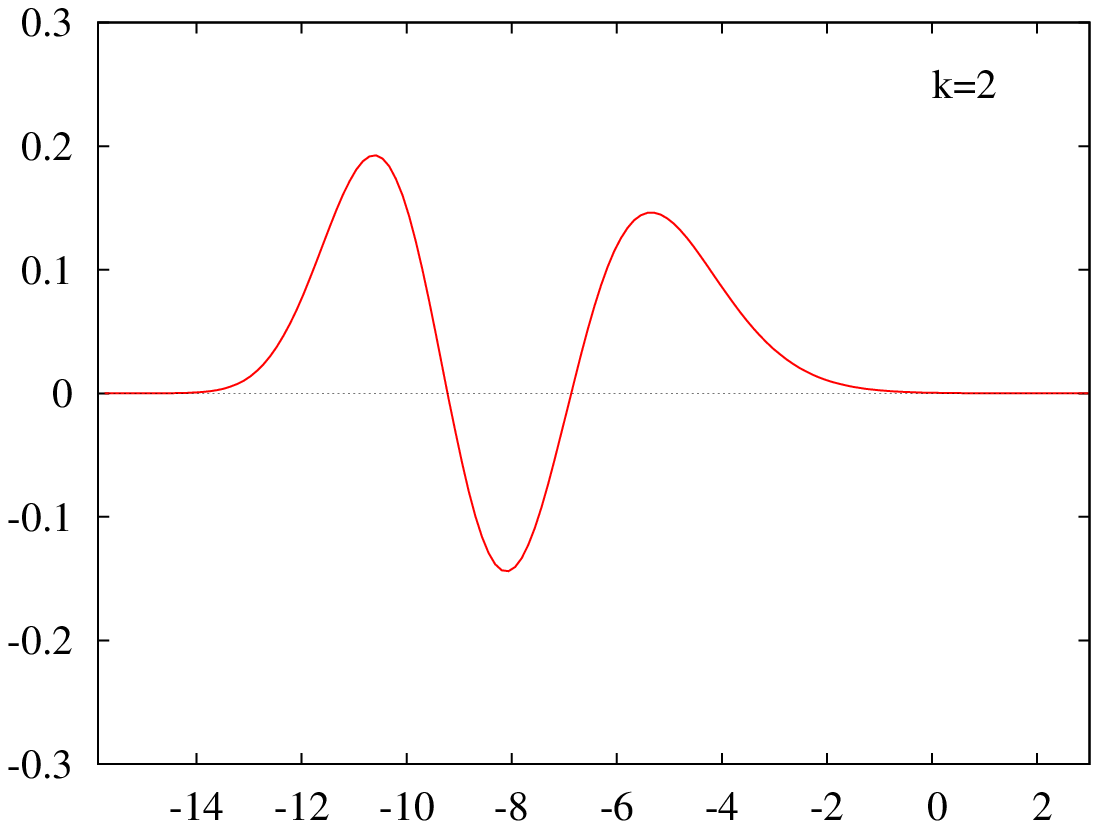}
\includegraphics[width=0.3\textwidth]{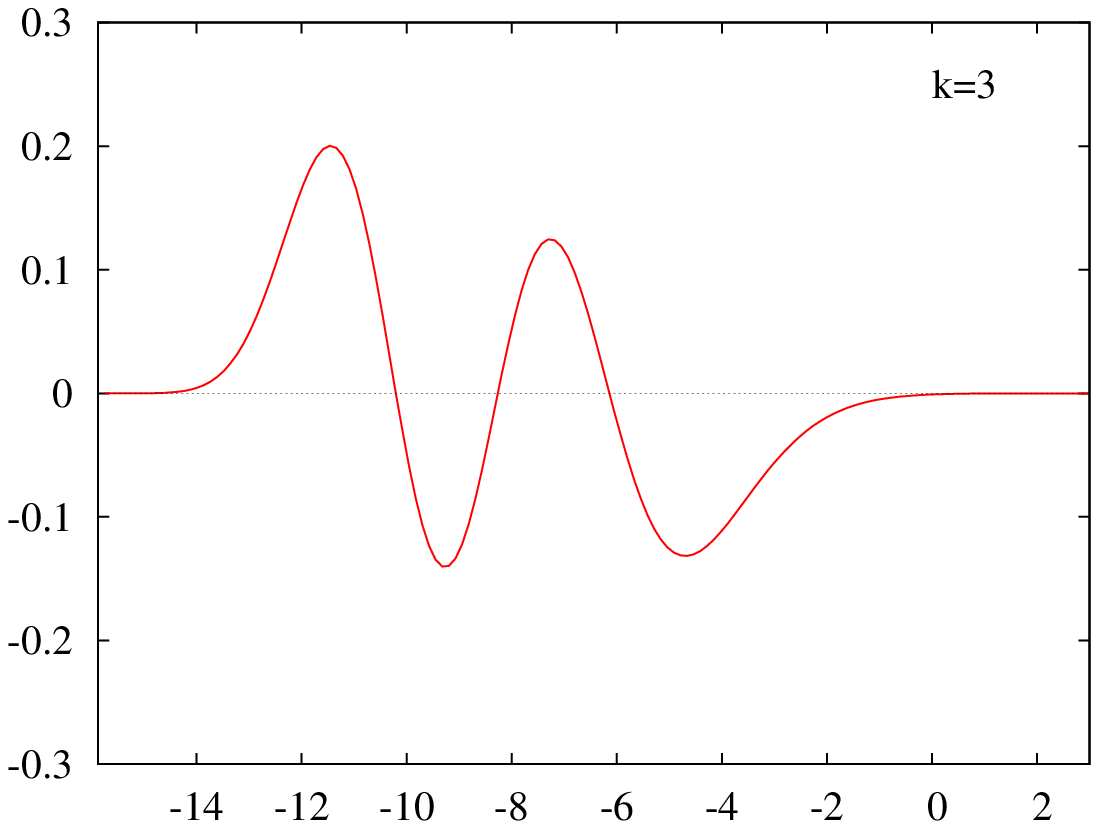}
\includegraphics[width=0.3\textwidth]{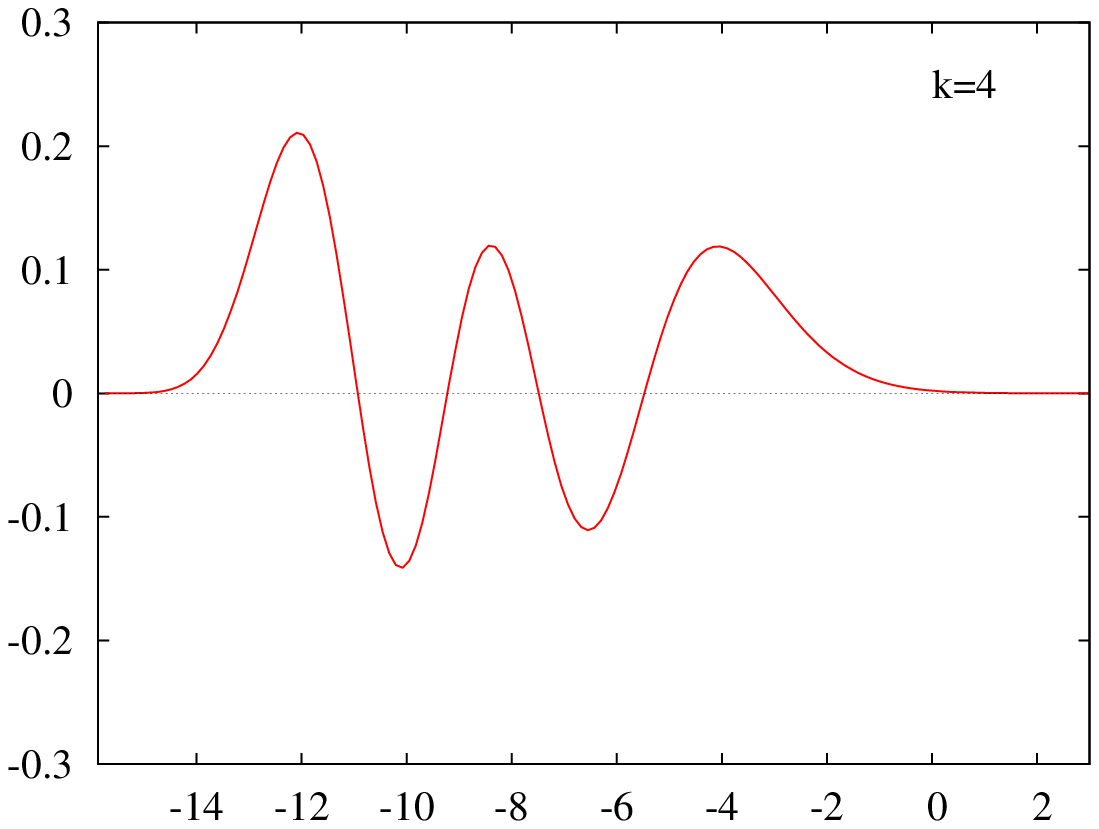}
\includegraphics[width=0.3\textwidth]{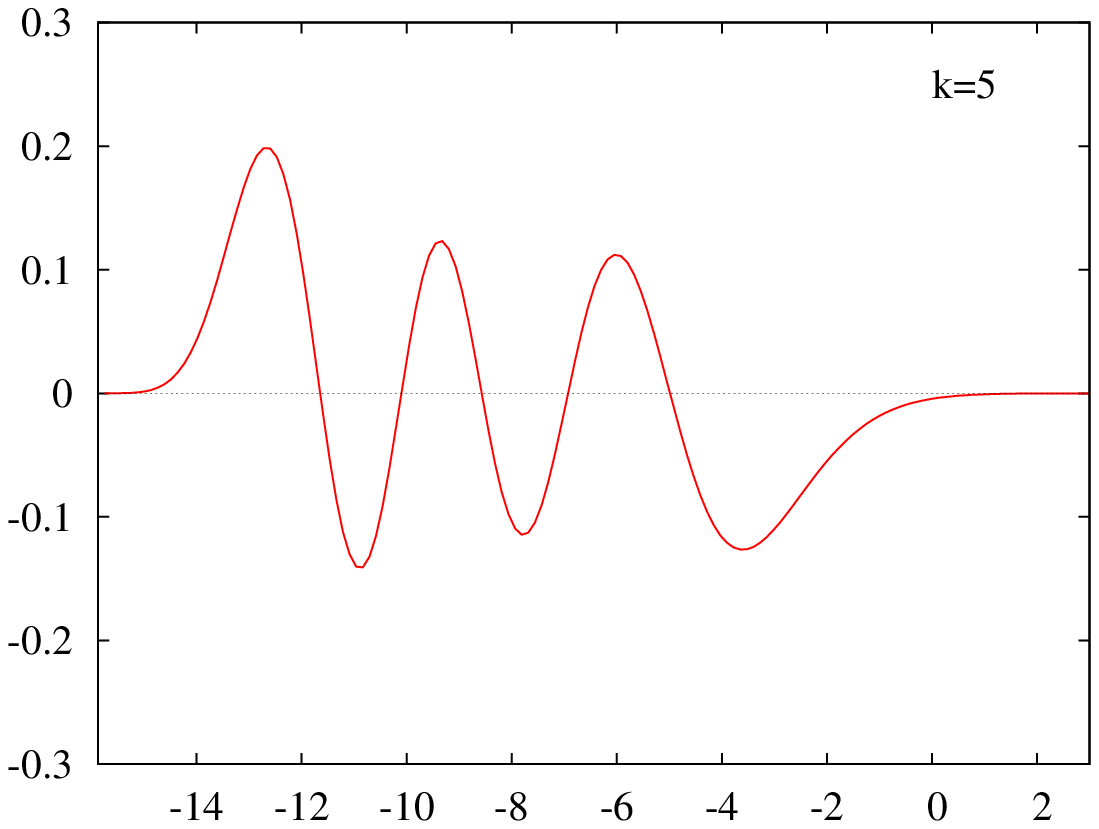}
\includegraphics[width=0.3\textwidth]{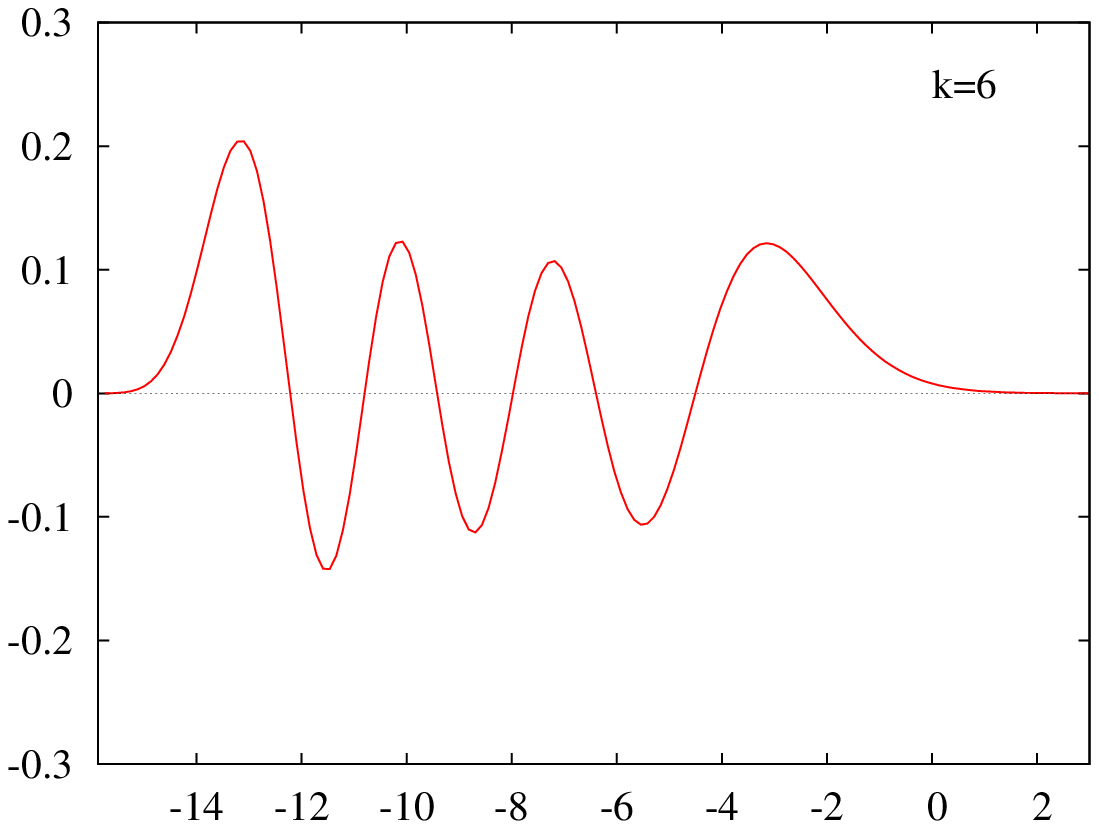}
\includegraphics[width=0.3\textwidth]{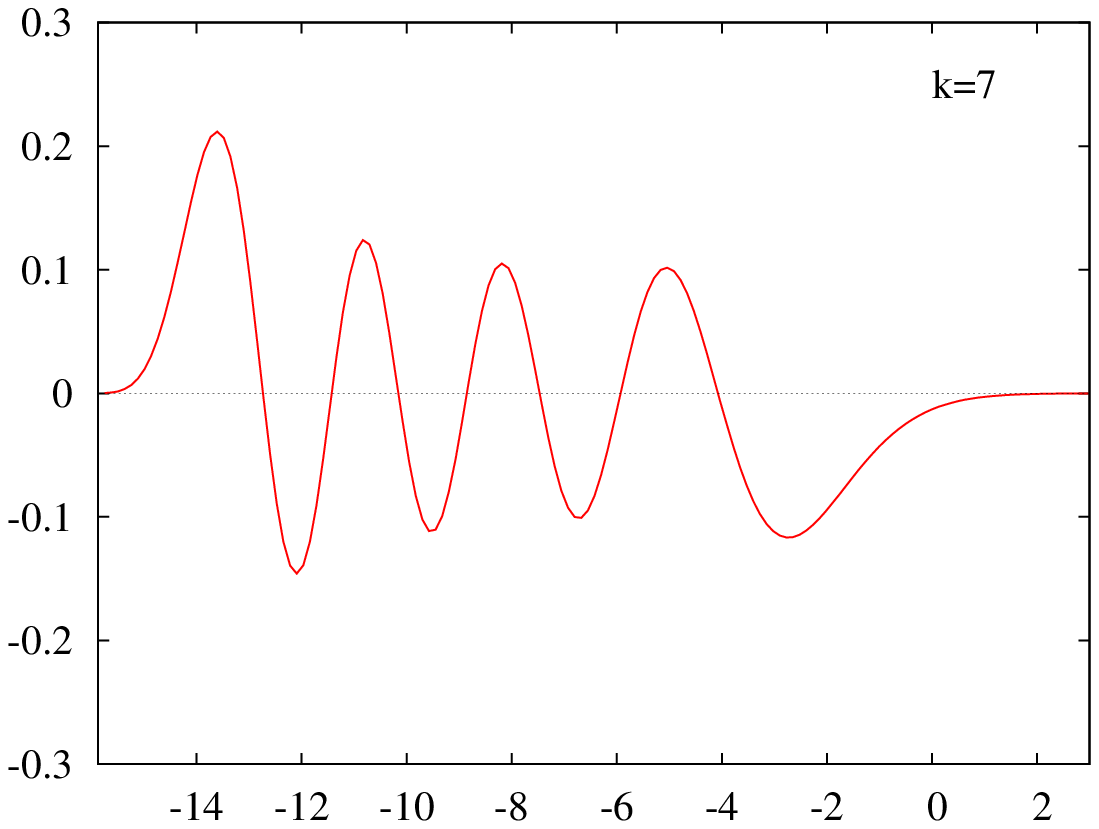}
\includegraphics[width=0.3\textwidth]{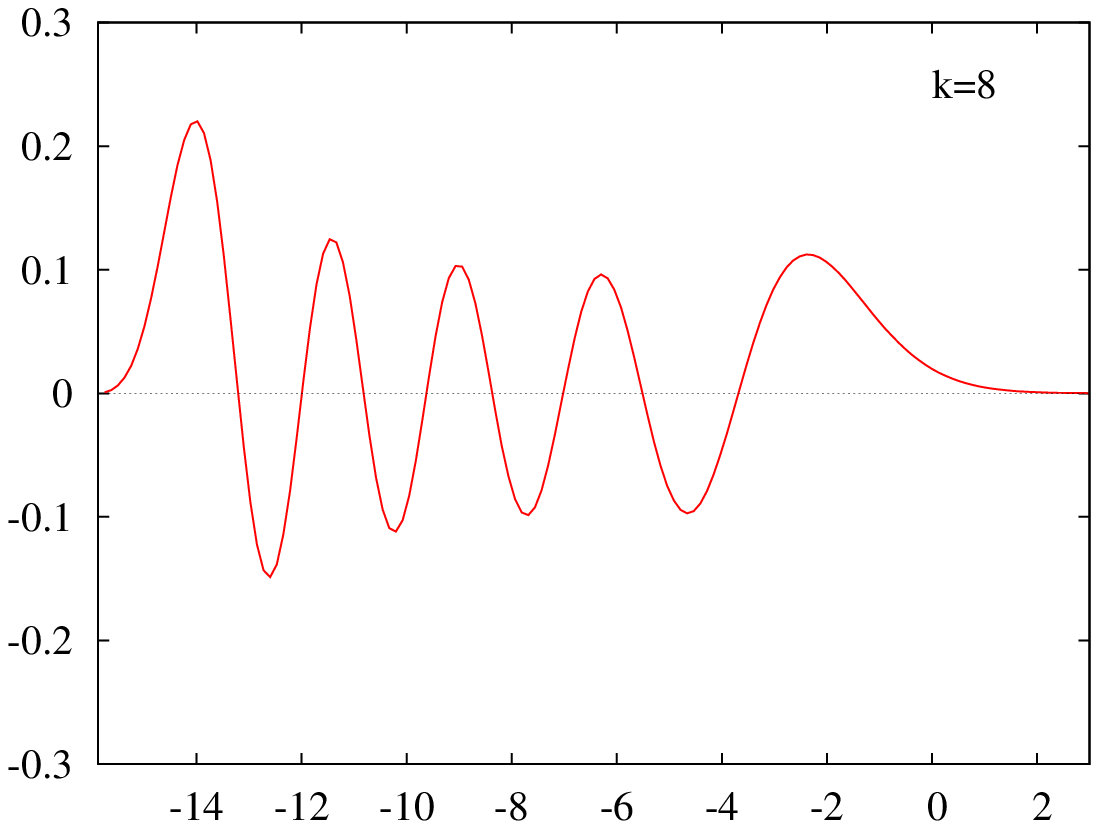}
\includegraphics[width=0.3\textwidth]{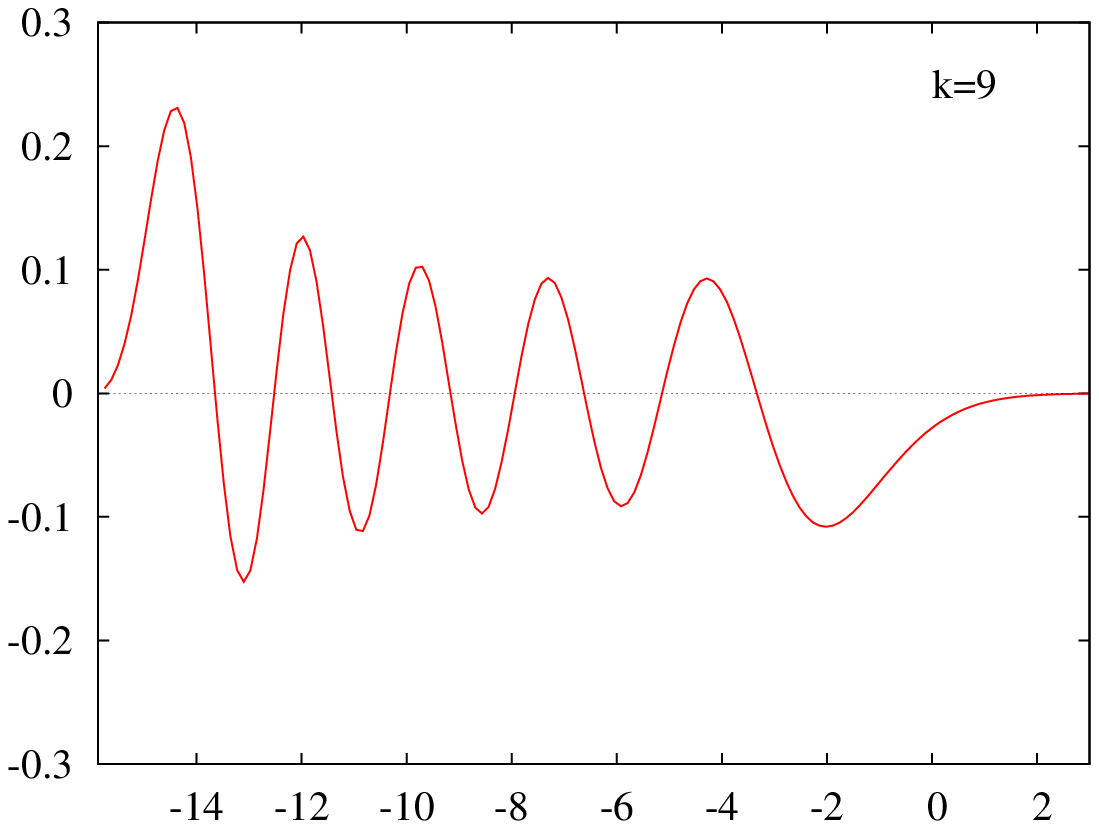}
\includegraphics[width=0.3\textwidth]{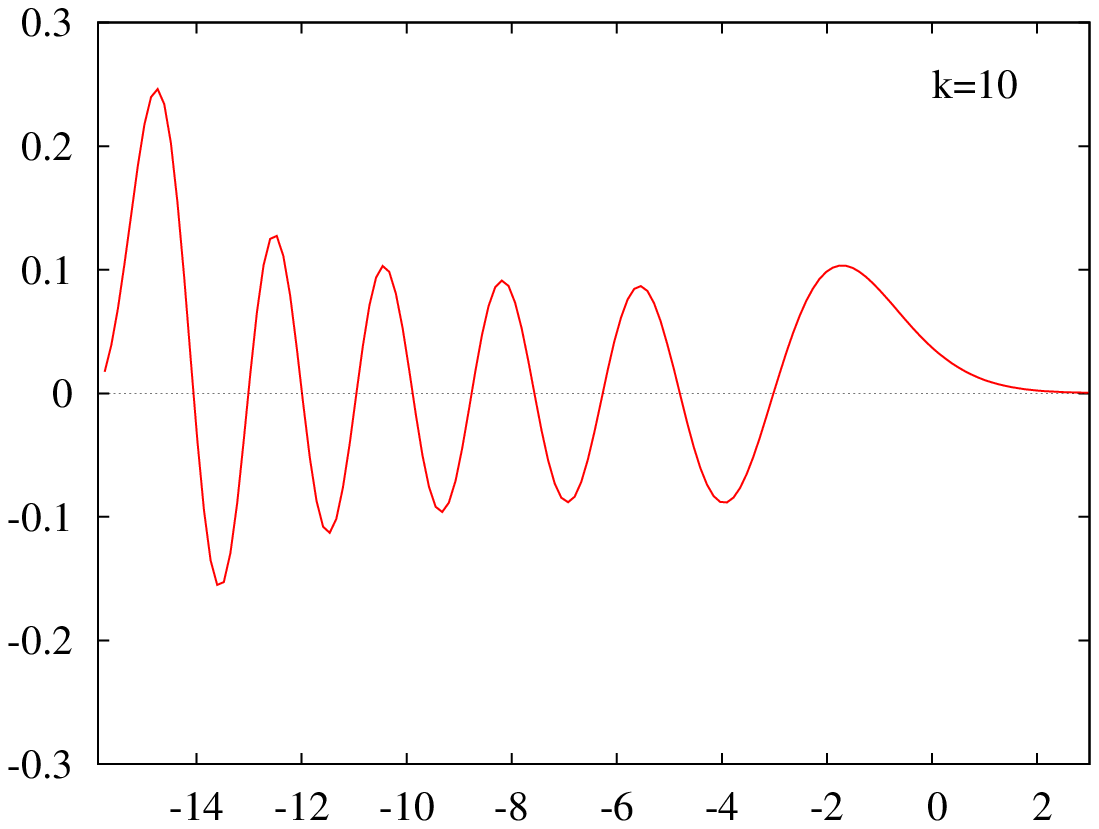}
\includegraphics[width=0.3\textwidth]{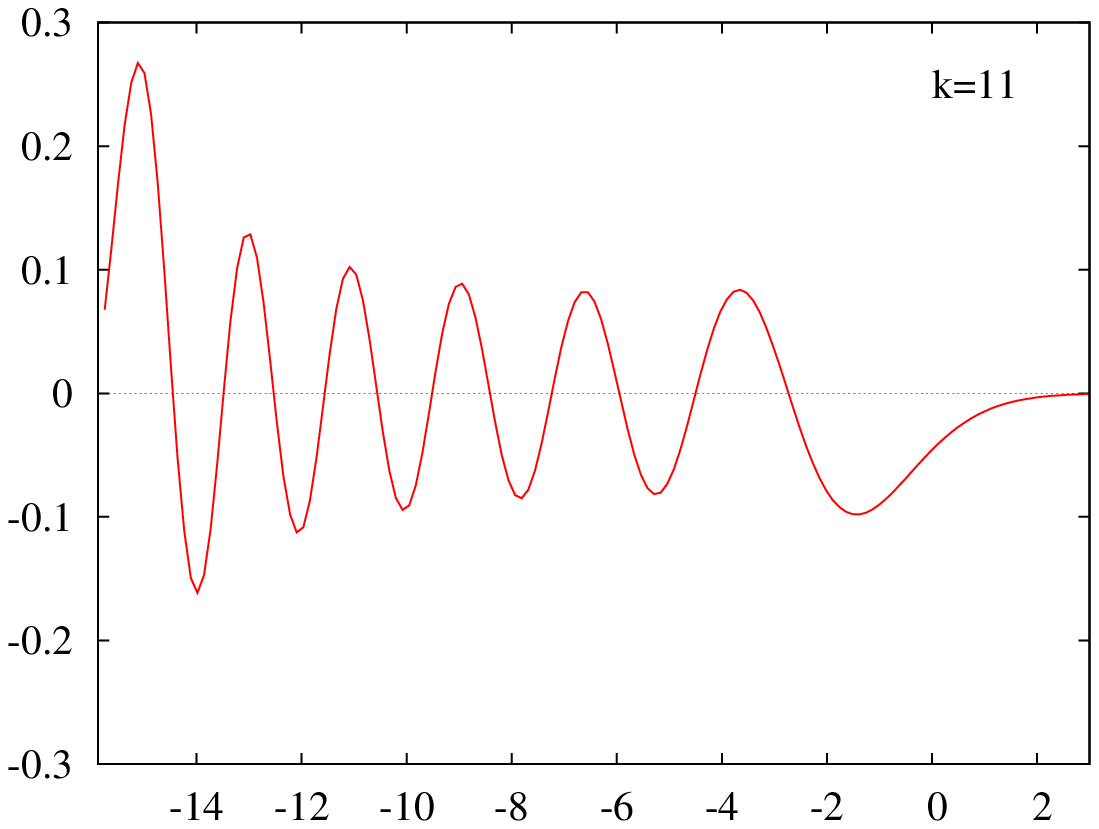}
\includegraphics[width=0.3\textwidth]{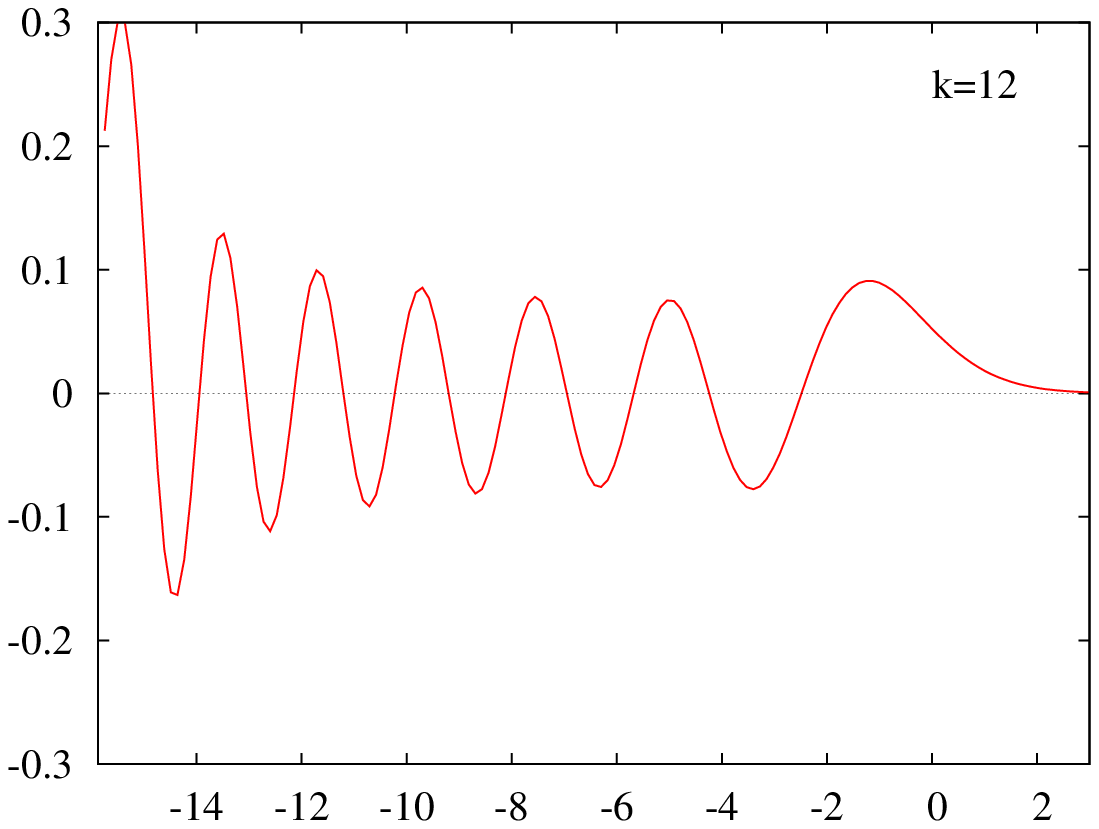}
\includegraphics[width=0.3\textwidth]{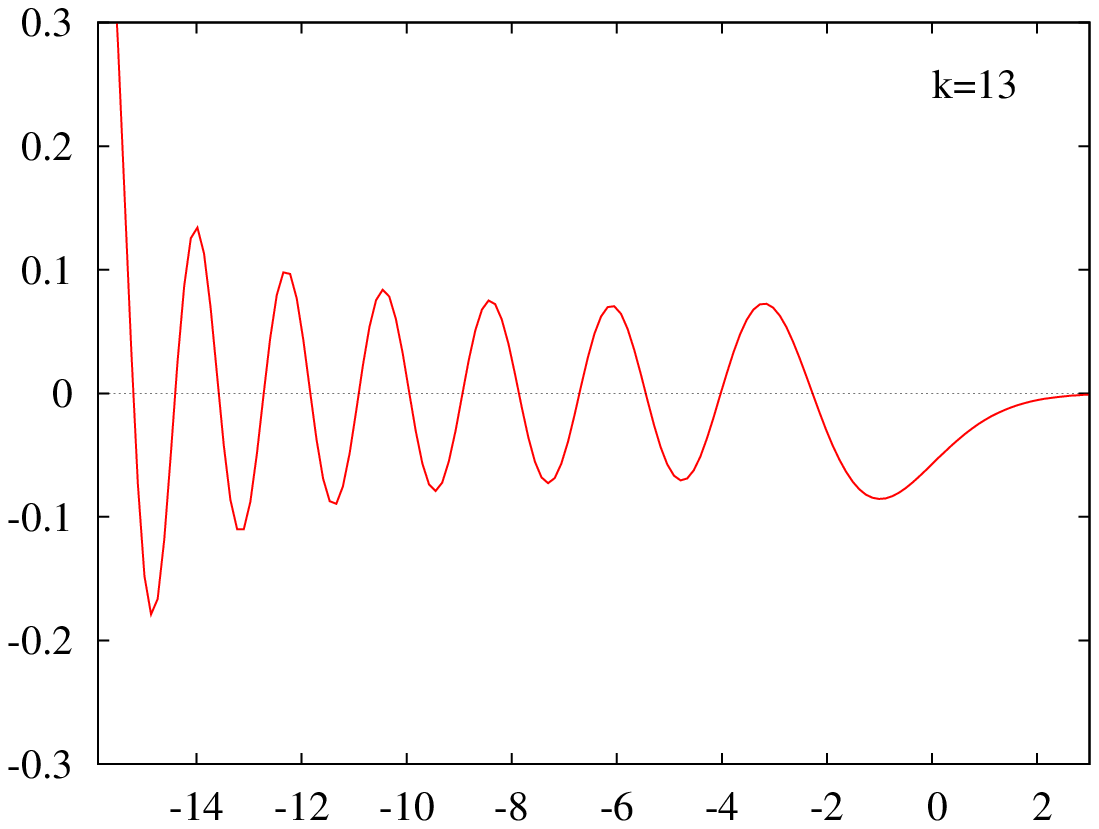}
\includegraphics[width=0.3\textwidth]{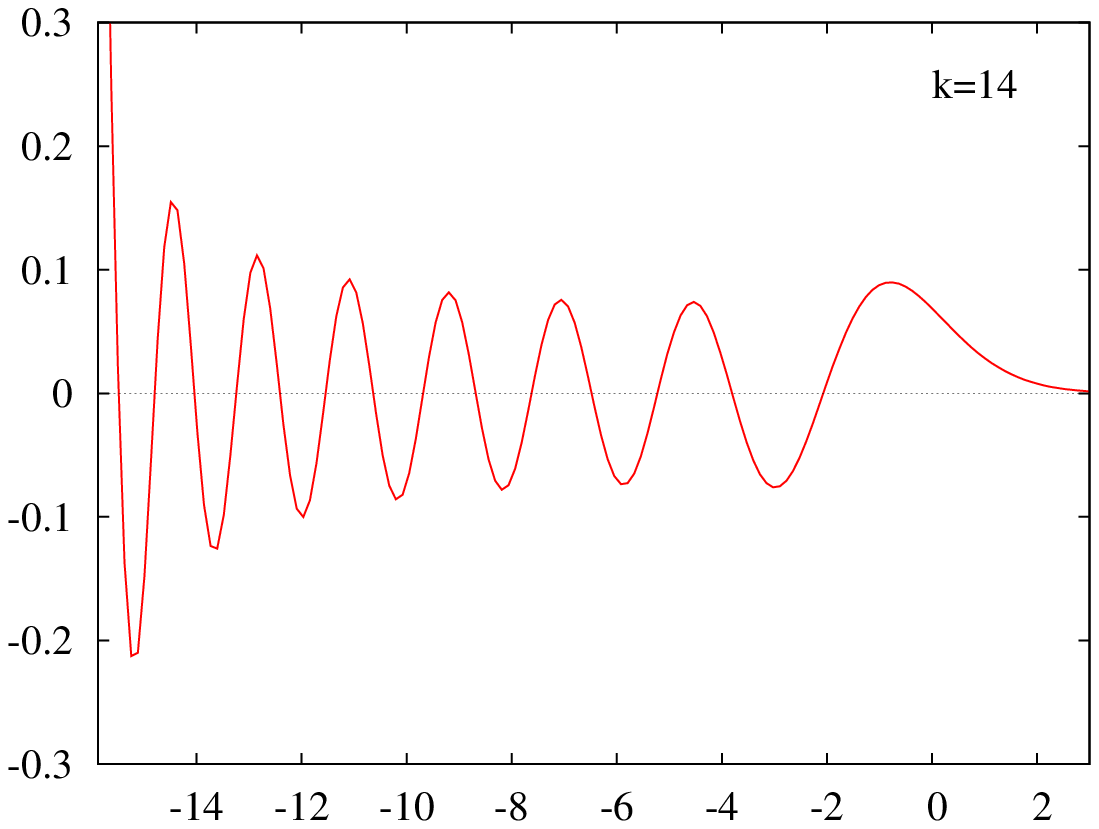}
\caption{Lowest eigenfunctions of $\mathcal{H}$ and of the Airy kernel as functions of $x$ for $s=15.875$. 
The left end of the figure is the boundary of the subsystem.}
\label{fig:evec}
\end{figure}

The first 15 eigenvectors for $s=15.875$ are shown in Fig. \ref{fig:evec} as functions of the 
variable $x$. There are about 150 points (sites) in the interval, so they appear as continuous 
curves and thus represent the continuous eigenfunctions of the Airy kernel (\ref{airy2}). They show 
also very similar features as those of the sine kernel \cite{Eisler/Peschel13}. Thus the lowest 
one looks like a Gaussian centered at $-s/2$, and for the higher states one has an increasing number 
of oscillations. Also, the amplitudes are smallest in the interior, which is seen especially in the 
left part. However, the empty region on the right leads to an asymmetry and, in particular, to 
the decay resembling the behaviour of an Airy function.
%
%

These features of the eigenfunctions can be understood qualitatively if one uses the fact that
the Airy kernel (\ref{airy2}) in the interval $(-s,\infty)$ commutes with the second-order 
differential operator \cite{Tracy/Widom94}
\eq{
D = -\frac{\dd}{\dd x} (x+s) \frac{\dd}{\dd x} + x(x+s) \, .
\label{diffop1}}
Therefore both operators have common eigenfunctions. Now, for $x$ near $-s/2$, $D$
can be written approximately
\eq{
D = \frac{s}{2} \left[ -\frac{\dd^2}{\dd x^2} +\frac{2}{s}(x+\frac{s}{2})^2 -\frac{s}{2} \right] 
\label{diffop2}}
and is seen to be the Hamiltonian of a harmonic oscillator with frequency $\omega^2=2/s$
centered at $x=-s/2$. This explains the similarity of the low eigenfunctions, concentrated
near $-s/2$, to those of the oscillator. For large $x$, on the other hand, one can transform the
eigenvalue equation $D \varphi=\mu \varphi$ directly into the Airy equation.

\section{Surface region: entanglement and fluctuations}

\begin{figure}[htb]
\center
\includegraphics[scale=0.7]{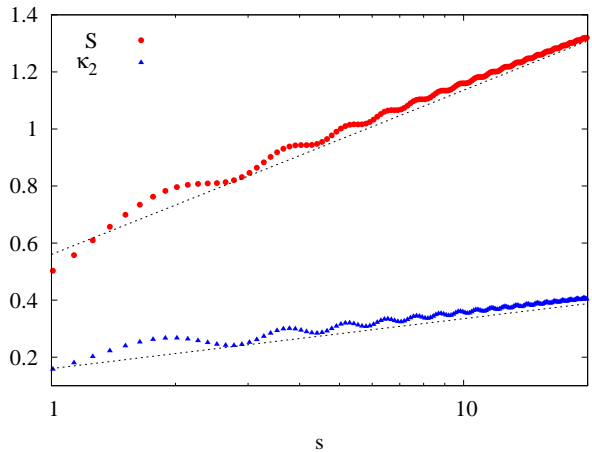}
\caption{Entanglement entropy (circles) and particle
number fluctuations (triangles) of the interval $(-s,\infty)$ as a function of $\ln s$ for 
$1<s<20$. The solid lines have slopes $1/4$ and $3/4\pi^2$, respectively.}
\label{fig:entfluct}
\end{figure}

Inserting the eigenvalues $\varepsilon_k$ into (\ref{entropy}) and (\ref{fluct}) gives the
results shown in Fig. \ref{fig:entfluct}, where $S$ and $\kappa_2$ are plotted against $\ln s$.
One sees a clear linear behaviour for larger values of $s$ with additional small oscillations 
which decrease with $s$. If one continues to higher $s$, the curves bend downwards since
one is leaving the surface region. However, one can follow the law further by increasing the
value of $\ell$. The lines in the figure have the slopes $1/4$ for the entropy and $3/4\pi^2$ for
the particle fluctuations and match the numerical results very well.

%
%

Thus one finds a logarithmic dependence on the subsystem size, but the prefactors differ
from the homogeneous case. In particular, $S$ is given by
\begin{equation}
S =  \frac{1}{4}\ln s + \mathrm{const} \, .
\label{entropy_surf}
\end{equation}
The ratio $S/\kappa_2 = \pi^2/3$, however, is the same as for homogeneous systems, where it 
follows from the behaviour of the cumulants in the full counting statistics 
\cite{Klich/Levitov09,Song11,Calabrese/Mintchev/Vicari12b}. In terms of the $\varepsilon_k$, 
it is a simple 
consequence of the approximately linear spectrum, which gives a constant density of states $N(0)$ 
if one changes the sums over $k$ into integrals over $\varepsilon_k$. Then (\ref{entropy}) and
(\ref{fluct}) become, with $\omega=\varepsilon_k/2$ and partial integrations in $S$,
\begin{equation}
S =  4N(0)\int_{0}^{\infty} d\omega \frac{\omega^2}{\ch^2\omega} = N(0) \frac{\pi^2}{3}
\label{entropy_cont}
\end{equation}
and 
\begin{equation}
\kappa_2 = N(0) \int_{0}^{\infty} d\omega \frac{1}{\ch^2\omega} = N(0)
\label{fluct_cont}
\end{equation}
from which the given ratio follows. 

\begin{figure}[htb]
\center
\includegraphics[scale=0.7]{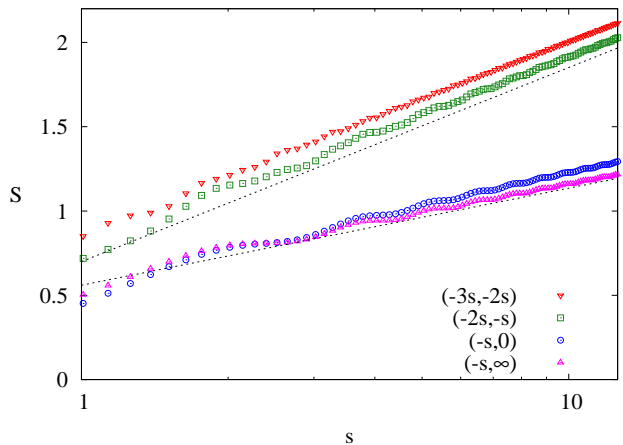}
\caption{Entanglement entropy as a function of $\ln s$ for subsystems of size $s$ at various 
positions. For comparison, also the semi-infinite case (Fig. 3) is shown.
The dotted black lines have slopes $1/2$ and $1/4$, respectively.}
\label{fig:entk}
\end{figure}

If the subsystem does not extend to infinity, but is an interval of length $s$ with both
end points in the surface region, one finds the results in Fig. \ref{fig:entk}. The entropy
varies again logarithmically with $s$, but there are two cases. If the right end is at $x=0$
($m=\ell$), the slope is again $1/4$ and the values are very close to those for the semi-infinite
case, which is also shown for comparison. If, however, the right end point is at a distance 
of order $s$ inside the surface, the slope lies close to $1/2$, i.e. it doubles. This is
again the same situation as for homogeneous systems, where the number of contact points
between subsystem and remainder enters into the prefactor of the logarithm. For the
fluctuations, one finds the same features, and the ratio $S/\kappa_2$ has again the value 
$\pi^2/3$.

%
%

These findings can be compared with results obtained by Soshnikov \cite{Soshnikov00} 
for the Airy kernel $K$. In his approach, one first introduces the scaling 
variables
\eq{
z_i = \frac{2}{3\pi} x_i |x_i|^{1/2}
\label{rescaling}}
in terms of which a new kernel can be defined as
\eq{
Q(z_1,z_2) = \frac{\pi}{|x_1|^{1/4}|x_2|^{1/4}} K(x_1,x_2) \, .
\label{defQ}}
The factor in front of $K(x_1,x_2)$ takes into account the change of variables and ensures 
$\Tr Q^n = \Tr K^n$ for arbitrary $n$. Now, one can use the series approximation of the Airy 
functions and expand $K(x_1,x_2)$ for negative arguments $x_1,x_2 < 0$. It turns out,
that there are only two important contributions in the infinite series and
one can write $Q = Q_1 + Q_2 + \Delta Q$ where
%
\begin{eqnarray}
Q_1(z_1,z_2) &= \frac{\sin \pi(z_1-z_2)}{\pi(z_1-z_2)} \left[
\frac{(z_1^2)^{1/3} + (z_1 z_2)^{1/3} + (z_2^2)^{1/3}}{3(z_1 z_2)^{1/3}}\right] , \\
Q_2(z_1,z_2) &= \frac{\cos \pi(z_1+z_2)}{\pi(z_1+z_2)} \left[
\frac{(z_1^2)^{1/3} - (z_1 z_2)^{1/3} + (z_2^2)^{1/3}}{3(z_1 z_2)^{1/3}}\right] .
\end{eqnarray}
%
The fluctuations, given by $\Tr Q(1-Q)$, can then be evaluated by using only the kernels
$Q_1$ and $Q_2$ and further proving that the contributions from $\Delta Q$ vanish.

One can gain insight into this procedure by comparing numerically the rescaled
Airy kernel $Q$ with its approximation $Q_1+Q_2$. One then finds that both for the
diagonal and the non-diagonal parts of the kernels, the sum $Q_1+Q_2$ gives a very good 
approximation to $Q$ as long as the arguments $z_1$ and $z_2$ are not too close to zero. 
Therefore this approximation is perfectly reasonable if one wants to obtain the fluctuations 
within an interval $-L < z_i < -1 $, as done by Soshnikov. In this way one obtains
\eq{
\kappa_2 = \Tr Q(1-Q) = \frac{1}{2\pi^2} \ln L + \mathrm{const} =
\frac{3}{4\pi^2} \ln s + \mathrm{const}
\label{soshnikov}}
if one uses $L=2 s^{3/2}/3\pi$.
The calculation is sketched in the Appendix, since there is a small mistake in \cite{Soshnikov00} 
(in the quantity $I_3(u)$ on p.508) which leads to a prefactor $11/18\pi^2$ there instead of 
$1/2\pi^2$ in (\ref{soshnikov}). Contrary to the statement in \cite{Soshnikov00}, the calculation
also gives a doubled prefactor for the shifted intervals. From (\ref{trace1bulk}), one finds for
an interval $(-ks,-(k-1)s)$
\eq{
\kappa_2 = \frac{3}{2\pi^2} \left[ \ln s  + \ln (k^{3/2}-(k-1)^{3/2}) \right] + \mathrm{const}\,,
\,\, \,\, k > 1 
\label{kappa_interval}}
and similarly for $S$. The second term causes an upward shift of the curve with increasing $k$ 
which is clearly seen in Fig. \ref{fig:entk}. 

Thus one obtains the numerical result for both types of subsystems in this way. The approach is 
also very instructive, because in the variables $z_i$ one comes back essentially to a homogeneous
problem. The kernel $Q_1$ is a sine kernel times a factor and $Q_2$ resembles a reflected sine 
kernel times a factor, but these factors do not influence the leading behaviour
and the final result in terms of the variable $L$ actually \emph{is} the 
homogeneous one. Thus it is the rescaling (\ref{rescaling}) which is responsible for the changed 
prefactors.

%

\section{Bulk region}

The scaling laws (\ref{entropy_surf}) and (\ref{soshnikov}) found in the previous section
describe the entanglement and particle fluctuations in the surface region.
However, if the distance $r$ of the subsystem boundary from the end of the interface
far exceeds the surface length scale $\ell_s$, one expects a crossover to the bulk behaviour.
In particular, if the boundary is in the center of the interface, $r=\ell$, the bulk scaling laws 
for $S$ and $\kappa_2$ are recovered \cite{Eisler/Igloi/Peschel09,AKR08}.

\begin{figure}[htb]
\center
\includegraphics[scale=0.61]{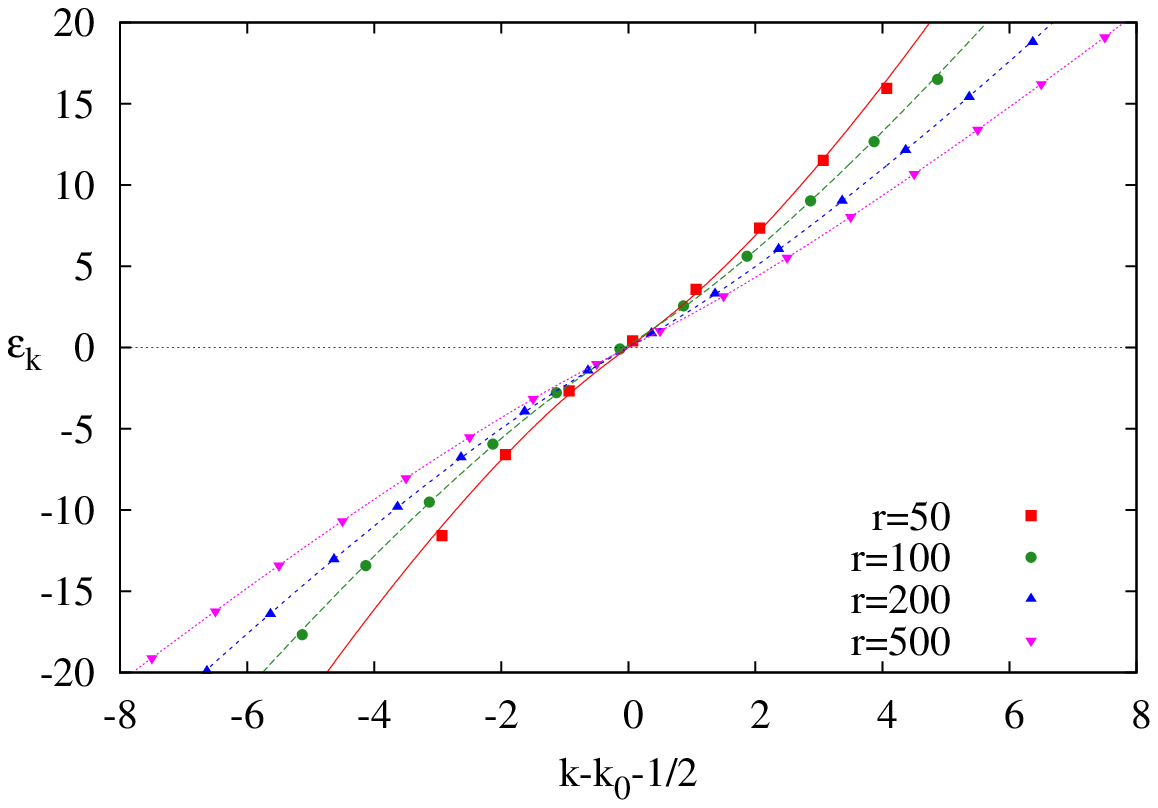}
\psfrag{L=500}[][][.6]{$\ell$=500}
\psfrag{L=1000}[][][.6]{$\ell$=1000}
\psfrag{c/L}[][][.6]{c/$\ell$}
\psfrag{z=r/L}[][][.6]{z=r/$\ell$}
\includegraphics[scale=0.61]{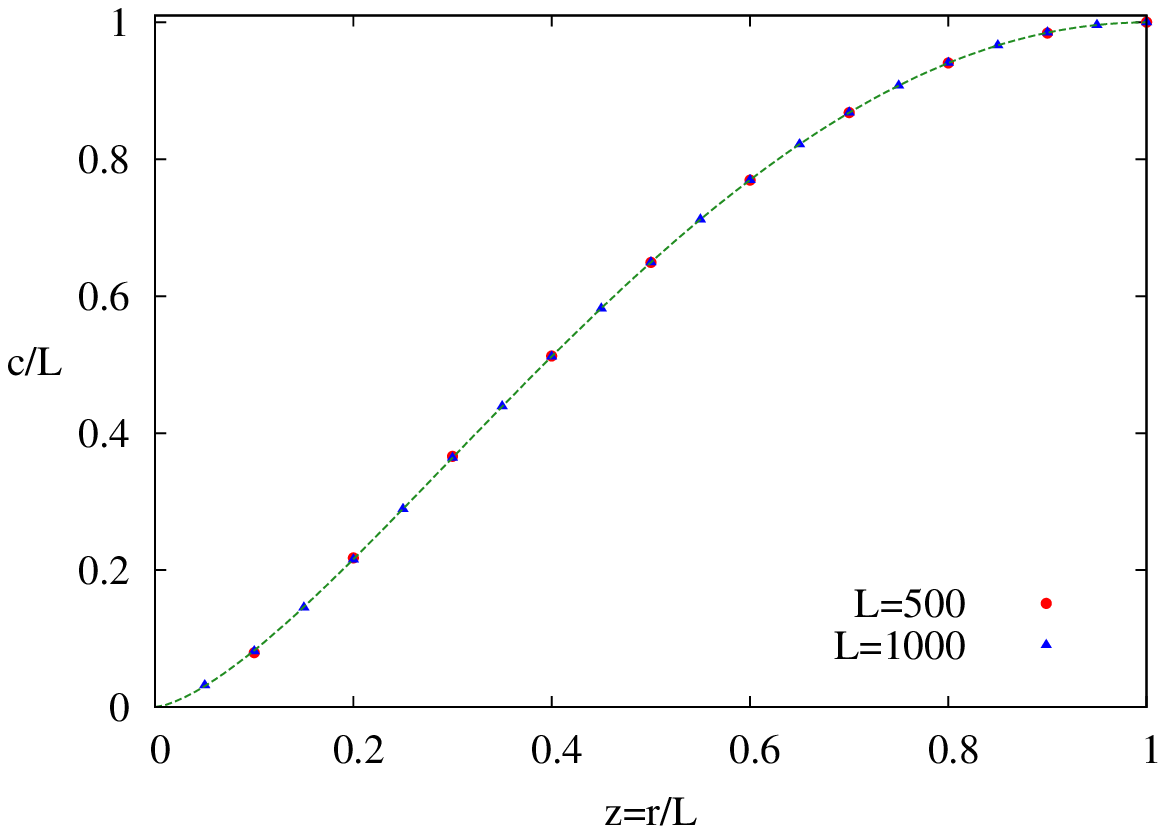}
\caption{Left: $\varepsilon_k$ for various distances $r$ of the subsystem boundary
from the edge of the interface with $\ell=1000$.
The lines are obtained by fitting Eq. (\ref{eq:slep}).
Right: Fitted values of $c/\ell$ plotted against $z=r/\ell$ for two different $\ell$.
The dashed line is the scaling function $\left[z(2-z)\right]^{3/2}$.}
\label{fig:slep}
\end{figure}

The crossover behaviour for intermediate values of $r$ is captured by the corresponding 
spectra $\varepsilon_k$. The initial filling up of the negative levels for $r \sim \ell_s$ 
(see Fig. \ref{fig:eps}) is followed by a slow decrease of the slope around $\varepsilon_k=0$, 
which persists for all values of $r \le \ell$. Shifting the spectra above each other, as shown 
on the left of Fig. \ref{fig:slep}, one obtains a similar plot as for the homogeneous chain, 
where the $\varepsilon_k$ follow from the eigenvalues of the sine kernel. In the latter case, 
the low-lying part of the spectrum is well described by the formula 
\cite{Eisler/Peschel13,Slepian65,Ivanov/Abanov13}
\eq{
\frac{\varepsilon_k}{2\pi} \ln 4c -
\varphi \left( \frac{\varepsilon_k}{2\pi}\right) =
\pi \left( k - k_0-\frac{1}{2} \right)
\label{eq:slep}}
with $\varphi (z)=\arg \Gamma (1/2 + iz)$. Note, that the factor $2$ difference
on the right hand side with respect to Refs. \cite{Eisler/Peschel13,Slepian65}
is due to the semi-infinite geometry, with a single boundary.

One can now view Eq. (\ref{eq:slep}) as an ansatz for the inverse function $k(\varepsilon)$
with the fitting parameters $k_0$ and $c$. The fitted curves, shown by the lines on
the left of Fig. \ref{fig:slep}, give a very good approximation for $\varepsilon_k$.
The parameter $k_0 $ is found to be equal to the average number of particles $\bar N$ in the
subsystem, in agreement with the homogeneous case \cite{Eisler/Peschel13} .
The parameter $c$ is obtained in a form $c=\ell f(z)$ with scaling variable $z=r/\ell$ and is
shown on the right of Fig. \ref{fig:slep}.
Due to the symmetry of the problem, the spectra for $r$ and $2\ell - r$ are identical and thus
the scaling function has the property $f(2-z)=f(z)$. Furthermore, matching the modified prefactor in the
surface regime $r \sim \ell_s$ requires $f(z) \sim z^{3/2}$ for $z \to 0$. Remarkably, the simplest choice
with these properties $f(z)=\left[ z(2-z) \right]^{3/2}$, shown by the dashed line in the right of 
Fig. \ref{fig:slep}, gives an excellent agreement with the numerical data.

%
%

\begin{figure}[htb]
\center
\psfrag{L=150}[][][.6]{$\ell$=150}
\psfrag{L=300}[][][.6]{$\ell$=300}
\psfrag{L=450}[][][.6]{$\ell$=450}
\psfrag{R=Lf(r/L)}[][][.6]{R=$\ell \,$f(r/$\ell$)}
\includegraphics[scale=0.61]{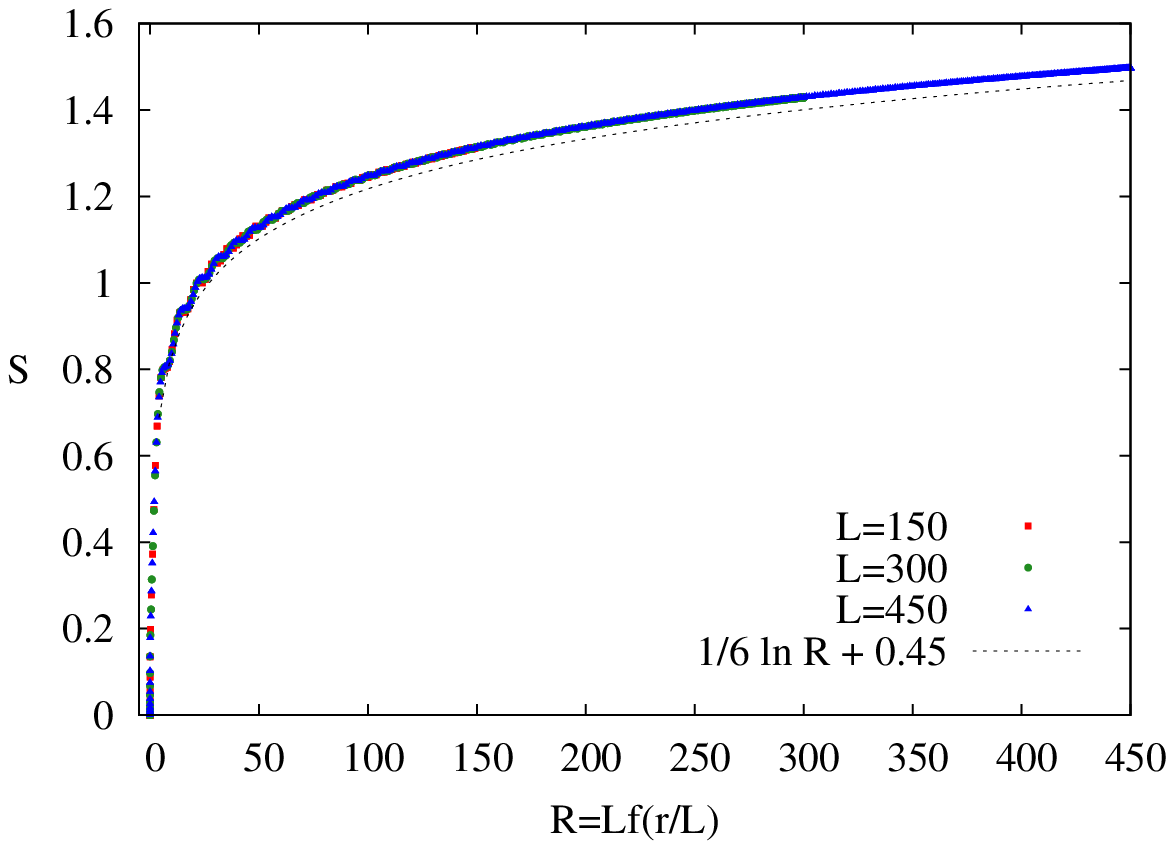}
\psfrag{L=150}[][][.6]{$\ell$=150}
\psfrag{L=300}[][][.6]{$\ell$=300}
\psfrag{L=450}[][][.6]{$\ell$=450}
\psfrag{R=Lf(r/L)}[][][.6]{R=$\ell \,$f(r/$\ell$)}
\includegraphics[scale=0.61]{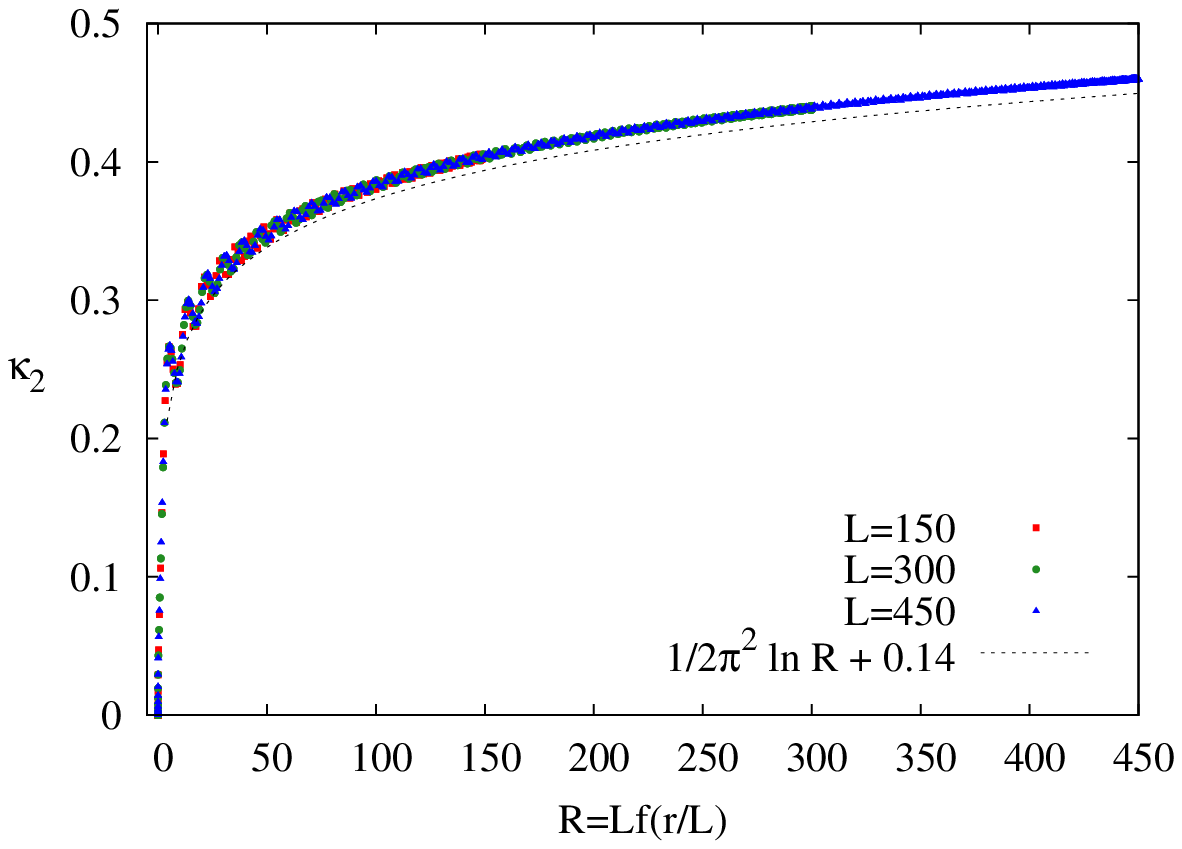}
\caption{Scaled entropy (left) and particle number fluctuations (right) as a fuction of the effective
length $R$ and for various $\ell$. The dotted lines are shown for comparison with the scaling functions.}
\label{fig:skscaled}
\end{figure}

It follows immediately from the above findings that the entropy has the form
\eq{
S = \frac{1}{6} \ln \left[ \ell f(r/\ell) \right] + \mathrm{const}
\label{eq:entsc}}
and a similar formula with prefactor $1/2\pi^2$ holds for $\kappa_2$. The effective
length $R=\ell f(r/\ell)$ appearing in the argument gives the correct limiting
cases $R \approx 2s^{3/2}$ for the surface region $r=\ell_s \, s$ and $R=\ell$ for 
the center of the interface $r=\ell$. The functions $S$ and $\kappa_2$ are plotted in Fig.
\ref{fig:skscaled} for the full range of $R\le\ell$ where one sees a perfect collapse for different $\ell$.
On top of the logarithmic growth, both feature oscillations which, however, decrease as one
moves deep into the bulk. In fact, $c = R$ holds only approximately, up to small oscillating
corrections. However, neglecting the deviations, the spectrum in Eq. (\ref{eq:slep})
is identical to that of a segment of length $R$ in a half-filled semi-infinite chain.
Thus, one can even infer the subleading constants $0.478$ for $S$
\cite{Ivanov/Abanov13,Jin/Korepin04} and $0.150$ for $\kappa_2$ \cite{ELR06,Song12}.
These predictions are shown slightly shifted by the dotted lines in Fig. \ref{fig:skscaled}
to allow for comparison.

%
%

Finally, one should mention that a similar scaling behaviour is expected for Fermi gases in a trap.
Indeed, in the surface region of a potential well the Airy kernel is obtained after appropriate rescaling of
the correlation kernel \cite{Eisler13}.
Furthermore, the bulk scaling behaviour of $S$ has also been observed for harmonic traps \cite{Vicari12}.
In general, for potentials of the form $V_p(x)=x^p/p$ with $p$ even, one expects for the entropy
\eq{
S = \frac{1}{6} \ln \left[ N f_p(r/\ell_{p}) \right] + \mathrm{const}
\label{eq:enttrap}}
where $N$ is the number of particles and $\ell_{p}=V^{-1}_{p}(E_N)$ is the classical turning point
for the particle at the Fermi level $E_N$ ($2\ell_p$ is the size of the density profile in the trap).
The scaling function must behave as $f_p(z) \sim z^{3/2}$
for $z \to 0$ in order to reproduce the surface scaling law in the region
$r \sim \ell_{s,p} = \left[ 2V'(\ell_p) \right]^{-1/3}$ \cite{Eisler13}.
In particular, we find for the harmonic trap $f_2(z)=f(z)$ which suggests a closer connection
to the problem with the linear potential. Note, however, that in general $f_p(z)$ depends on the exponent $p$
and for $p \to \infty$ one must recover $f_{\infty}=\sin (\pi z/2)$ which is the
scaling function for the hard-wall trap \cite{CMV11a,CMV11b}.
In this case one has $f_{\infty}(z) \sim z$ for $z \to 0$, since the surface region vanishes completely.
Thus the transition from the soft-wall to the hard-wall trap must, in some way, be singular, the details
of which are yet to be uncovered.

\section{Summary}

We have studied the entanglement in a fermionic chain where the particle density varies and a
surface region with an own length scale exists. In this surface region, we determined the nature of 
the entanglement Hamiltonian by looking at its single-particle eigenvalues and eigenfunctions.
We found not only the 
well-known spectra but also the usual logarithmic behaviour of the entanglement 
entropy. The conformal prefactor, however, turned out to be modified by a factor of 3/2, and 
one could interpret this as an effective central charge $c_{\mathrm{eff}}=3/2\,c$. 

Such effective 
central charges have been found also in other situations. For example, a defect in a free-particle 
chain leads to a continuously varying $c_{\mathrm{eff}}$
\cite{Peschel05,Igloi/Szatmari/Lin09,Eisler/Peschel10,Peschel/Eisler12,Calabrese/Mintchev/Vicari12a}
and in strongly random spin chains one finds 
$c_{\mathrm{eff}}=c\, \ln2$ \cite{Refael/Moore09}. Also a trap in a critical XY chain modifies $c$,
if one works with the true trap size, namely by a factor $p/(p+1)$, if the potential varies 
as $|n|^p$ \cite{CV10a}. However, in this case there is no well-defined surface and the effect 
results from a bulk length scale. By contrast, we have both bulk and surface regions and can 
sample both. Finally, while in all these examples $c$ is reduced, there are
also special random or aperiodic systems where it is enhanced as in our case
\cite{Santachiara06,Binosi/etal07,Hoyos/etal11,Igloi/Juhasz/Zimboras07}.

It is tempting to connect our result with the variation of the density in the surface. In the 
variable $x$, it is given by $\rho(x)= K(x,x)= \Ai'(x)^2-x\Ai(x)^2$ and varies roughly as $(-x)^{1/2}$. 
Integrating it in the interval $(-s,0)$, or also in shifted intervals, gives a particle number 
$\sim s^{3/2}$. However, the particle number has no symmetry with respect to the centre of the 
interface and one cannot arrive at the scaling function found in Section 5 in this way.

On the analytical side, we only invoked fluctuation results for the Airy kernel. However,
using the commuting differential operator, one probably can obtain also expressions for
the low-lying eigenvalues $\varepsilon_k$ in the large-$s$ limit. This would be analogous to 
the procedure used for the sine kernel in \cite{Slepian65,DesCloizeaux/Mehta73}, see also
\cite{Eisler/Peschel13}. Such a calculation was started in \cite{Tracy/Widom94}, but limited 
to $\zeta_k$ in the vicinity of one, whereas small $\varepsilon_k$ correspond to $\zeta_k$
near one-half.


\ack{
The work of V.E. was realized in the framework of T\'AMOP 4.2.4.A/1-11-1-2012-0001
``National Excellence Program''. The project was supported by the European Union
and co-financed by the European Social Fund. We thank Ferenc Igl\'oi for a critical reading of
the manuscript. V.E. also thanks for the hospitality of 
Freie Universit\"at Berlin where part of this work was completed.}


\section*{Appendix}

We calculate here the particle fluctuations in the subsystem, using 
$\kappa_2 = \Tr Q(1-Q)$ with $Q=Q_1+Q_2$ and following \cite{Soshnikov00}. The main qantities are 
$\Tr Q_1^2$ and $\Tr Q_2^2$.

We first consider the interval $-L \le z_i \le -1 $. The trace then denotes
\eq{
\Tr AB = \int_{-L}^{-1} \int_{-L}^{-1}  A(z_1,z_2) B(z_2,z_1) \dd z_1 \dd z_2 \, .
}
For $\Tr Q_1^2$ one introduces new variables $u=z_1-z_2$ and $z=z_2$ to rewrite the 
integral as
\eq{
\Tr Q_1^2 = \frac{2}{9} \int_{0}^{L-1} \left( \frac{\sin \pi u}{\pi u} \right)^{2} I_1(u)du
\label{traceQ1}
}
where 
\eq{
I_1(u) = \int_{1}^{L-u} \left[ \left( \frac{z+u}{z} \right)^{1/3} + 1 +
\left(\frac{z}{z+u} \right)^{1/3} \right]^2 \dd z \, .
\label{integral1indef}
}
This integral can be calculated in closed form 
\eq{
I_1(u) = 3 \left[ z+ z^{1/3}(z+u)^{2/3}+ z^{2/3}(z+u)^{1/3} \right]_{z=1}^{z=L-u} 
}
which yields
\eq{
\fl
I_1(u) = 3 \left[ (L-u-1) + L \left( 1 - \frac{u}{L} \right)^{2/3} + L \left( 1 - \frac{u}{L} \right)^{1/3}
- (1+u)^{2/3} - (1+u)^{1/3} \right]
}
For $u \lesssim L$, this can be approximated by $I_1(u) \simeq 9L-6u$. Inserting this into (\ref{traceQ1}), 
the first term gives $L$, which is cancelled by $\Tr Q_1 = L-1$, and the second term leads to a logarithm
due to the $1/u$ behaviour of the integrand. The result is 
\eq{
\Tr Q_1(1-Q_1) = \frac{2}{3\pi^2} \ln L + \mathrm{const} \, .
\label{trace1}}
The quantity $\Tr Q_2^2$ can be evaluated in a similar way. In this case one has
\eq{
\Tr Q_2^2 = \frac{1}{9} \left[
\int_{2}^{L+1} \left( \frac{\cos \pi u}{\pi u} \right)^{2} I_2(u) \, \dd u +
\int_{L+1}^{2L} \left( \frac{\cos \pi u}{\pi u} \right)^{2} I_3(u) \, \dd u \right]
\label{traceQ2}
}
with the exact expressions
%
\begin{eqnarray}
I_2(u) &= \int_{1}^{u-1} \left[ \left( \frac{z}{u-z} \right)^{1/3} - 1 +
\left(\frac{u-z}{z} \right)^{1/3} \right]^2 \dd z \\ \nonumber
&= 3(u-2) - 6 (u-1)^{2/3} + 6 (u-1)^{1/3} \, , \\
I_3(u) &= \int_{u-L}^{L} \left[ \left( \frac{z}{u-z} \right)^{1/3} - 1 +
\left(\frac{u-z}{z} \right)^{1/3} \right]^2 \dd z \\ \nonumber
&= 3(2L-u) + 6L \left(\frac{u}{L}-1\right)^{2/3} - 6L \left(\frac{u}{L}-1\right)^{1/3} .
\end{eqnarray}
%
Due to the integration limits, $I_3(u)$ gives only a finite contribution for $L\to \infty$.
The leading contribution from  $I_2(u)$ is obtained by putting $I_2(u) \simeq 3u$. Inserting this
into  (\ref{traceQ2}), one finds
\eq{
\Tr Q_2(1-Q_2) = -\frac{1}{6\pi^2} \ln L + \mathrm{const}
\label{trace2}}
since here $\Tr Q_2$ is finite for large $L$.  The same can be shown for $\Tr Q_1 Q_2$.
Thus adding (\ref{trace1}) and (\ref{trace2}), one obtains the result
\eq{
\Tr Q(1-Q) = \frac{1}{2\pi^2} \ln L + \mathrm{const}
\label{traceQ}}
given in (\ref{soshnikov}).

It is easy to see, what the changes are for intervals  $-L_2 \le z_i \le -L_1 $. The integration limits 
then change correspondingly and are $z=L_1$ and  $z=L_2-u$ in (\ref{integral1indef}). As a result, the 
approximate expression for $I_1(u)$ has contributions from all terms and becomes
$I_1(u) \simeq 9(L_2-L_1)-9u$. Inserting this into (\ref{traceQ1})
with upper limit $L_2-L_1$, one obtains for large $L_2-L_1$
\eq{
\Tr Q_1(1-Q_1) = \frac{1}{\pi^2} \ln(L_2-L_1) + \mathrm{const} \, .
\label{trace1bulk}}
There is no logarithmic contribution from $Q_2$ in this case, since all integration limits in 
(\ref{traceQ2}) are large. Therefore (\ref{trace1bulk}) is already the final result for $\kappa_2$
and the prefactor is twice as large as in (\ref{traceQ}).

\section*{References}



\begin{thebibliography}{99}


\bibitem{CC09}
Calabrese P and Cardy J 2009 {\em J. Phys. A: Math. Theor.\/} {\bf 42} 504005

\bibitem{Eisler/Igloi/Peschel09}
Eisler V, Igl\'oi F and Peschel I 2009 {\em J. Stat. Mech.\/}  P02011

\bibitem{CV10a}
Campostrini M and Vicari E 2010 {\em Phys. Rev. A\/} {\bf 81} 023606

\bibitem{CV10b}
Campostrini M and Vicari E 2010 {\em Phys. Rev. A\/} {\bf 81} 063614

\bibitem{CV10c}
Campostrini M and Vicari E 2010 {\em J. Stat. Mech.\/} P08020

\bibitem{Calabrese/Mintchev/Vicari12a}
Calabrese P, Mintchev M and Vicari E 2012 {\em  J. Phys. A: Math. Theor.\/} {\bf 45}
  105206

\bibitem{Vicari12}
Vicari E 2012 {\em Phys. Rev. A\/} {\bf 85} 062104

\bibitem{ARRS99}
Antal T, R{\'a}cz Z, R{\'a}kos A and Sch{\"u}tz G~M 1999 {\em Phys. Rev. E\/}
  {\bf 59} 4912

\bibitem{HRS04}
Hunyadi V, R{\'a}cz Z and Sasv{\'a}ri L 2004 {\em Phys. Rev. E\/}
{\bf 69} 066103

\bibitem{Sabetta/Misguich13}
Sabetta T and Misguich G 2013 {\em Phys. Rev. B\/} {\bf 88} 245114

\bibitem{Eisler/Racz13}
Eisler V and R\'acz Z 2013 {\em Phys. Rev. Lett.\/} {\bf 110} 060602

\bibitem{Eisler13}
Eisler V 2013 {\em Phys. Rev. Lett.\/} {\bf 111} 080402

\bibitem{Mehta04}
Mehta M~L 2004 {\em Random Matrices\/} 3rd ed (Amsterdam: Elsevier)

\bibitem{Soshnikov00}
Soshnikov A~B 2000 {\em J. Stat. Phys.\/} {\bf 100} 491


\bibitem{Peschel03}
Peschel I 2003 {\em J. Phys. A: Math. Gen.\/} {\bf 36} L205

\bibitem{Cheong/Henley04}
Cheong S~A and Henley C L 2004 {\em Phys. Rev. B\/} {\bf 69} 075111


\bibitem{Peschel/Eisler09}
Peschel I and Eisler V 2009 {\em J. Phys. A: Math. Theor.\/} {\bf 42} 504003

\bibitem{Eisler/Peschel13}
Eisler V and Peschel I 2013 {\em J. Stat. Mech.\/}  P04028

\bibitem{Tracy/Widom94}
Tracy C~A and Widom H 1994 {\em Commun. Math. Phys.\/} {\bf 159} 151


\bibitem{Klich/Levitov09}
Klich I and Levitov L 2009 {\em Phys. Rev. Lett.\/} {\bf 102} 100502

\bibitem{Song11}
Song H~F, Flindt C, Rachel S, Klich I and Le Hur K 2012 {\em  Phys. Rev. B\/} {\bf 83} 161408(R)

\bibitem{Calabrese/Mintchev/Vicari12b}
Calabrese P, Mintchev M and Vicari E 2012 {\em Europhys. Lett.\/} {\bf 98} 20003


\bibitem{AKR08}
Antal T, Krapivsky P~L and R\'akos A 2008 {\em Phys. Rev. E\/} {\bf 78} 061115

\bibitem{Slepian65}
Slepian D 1965 {\em J. Math. and Phys.\/} {\bf 44} 99

\bibitem{Ivanov/Abanov13}
Ivanov D~A and Abanov A~G 2013 {\em J. Phys. A: Math. Theor.\/} {\bf 46} 375005

\bibitem{Jin/Korepin04}
Jin B~Q and Korepin V~E 2004  {\em J. Stat. Phys.\/} {\bf 116} 79

\bibitem{ELR06}
Eisler V, Legeza {\"O} and R{\'a}cz Z 2006 {\em J. Stat. Mech.\/}  P11013

\bibitem{Song12}
Song H~F, Rachel S, Flindt C, Klich I, Laflorencie N and Le Hur K 2012 {\em
  Phys. Rev. B\/} {\bf 85} 035409

\bibitem{CMV11a}
Calabrese P, Mintchev M and Vicari E 2011 {\em Phys. Rev. Lett.\/} {\bf 107}
  020601

\bibitem{CMV11b}
Calabrese P, Mintchev M and Vicari E 2011 {\em J. Stat. Mech.\/}  P09028


\bibitem{Peschel05}
Peschel I 2005 {\em J. Phys. A: Math. Gen.\/} {\bf 38} 4327

\bibitem{Igloi/Szatmari/Lin09}
Igl\'oi F, Szatm\'ari Z and Lin Y-C 2009 {\em Phys. Rev. B\/} {\bf 80} 024405

\bibitem{Eisler/Peschel10}
Eisler V and Peschel I 2010 {\em Ann. Phys. (Berlin)\/} {\bf 522} 679

\bibitem{Peschel/Eisler12}
Peschel I and Eisler V 2012 {\em J. Phys. A: Math. Theor.\/} {\bf 45} 155301

\bibitem{Refael/Moore09}
Refael G and Moore J~E 2009 {\em J. Phys. A: Math. Theor.\/} {\bf 42} 504010

\bibitem{Santachiara06}
Santachiara R 2006 {\em J. Stat. Mech.\/} L06002

\bibitem{Binosi/etal07}
Binosi D, De Chiara G, Montangero S and Recati A 2007 {\em Phys. Rev. B\/} {\bf 76} 140405(R)

\bibitem{Hoyos/etal11}
Hoyos J~A, Laflorencie N, Viera A~P and Vojta T 2011 {\em Europhys. Lett.\/} {\bf 93} 30004

\bibitem{Igloi/Juhasz/Zimboras07}
Igl\'oi F, Juh\'asz R and Zimbor\'as Z 2007 {\em Europhys. Lett.\/} {\bf 79}, 37001

\bibitem{DesCloizeaux/Mehta73}
des Cloizeaux J and Mehta M~L 1973 {\em J. Math. Phys.\/} {\bf 13} 174

\end{thebibliography}


\end{document}